\documentclass[aps,prl,reprint,superscriptaddress,amsmath,amssymb]{revtex4-1}

\usepackage{graphicx}
\usepackage{adjustbox}
\usepackage{bm}
\usepackage{dcolumn}
\usepackage{multirow}
\usepackage{colortbl}
\usepackage{algorithm}
\usepackage[noend]{algpseudocode}
\usepackage{hyperref}
\hypersetup{colorlinks=true, citecolor=blue, urlcolor=blue, linkcolor=blue}

\begin{document}

\title{Predicting ground state configuration of energy landscape ensemble using graph neural network}

\author{Seong Ho Pahng}
 \email[]{spahng@g.harvard.edu}
 \affiliation{Department of Chemistry and Chemical Biology, Harvard University, Cambridge, MA 02138}
\author{Michael P. Brenner}
 \affiliation{School of Engineering and Applied Sciences, Harvard University, Cambridge, MA 02138}
 \affiliation{Google Research, Mountain View, CA 94043}

\date{\today}

\begin{abstract}
 Many scientific problems seek to find the ground state in a rugged energy landscape, a task that becomes prohibitively difficult for large systems. Within a particular class of problems, however, the short-range correlations within energy minima might be independent of system size. Can these correlations be inferred from small problems with known ground states to accelerate the search for the ground states of larger problems? Here, we demonstrate the strategy on Ising spin glasses, where the interaction matrices are drawn from protein contact maps. We use graph neural network to learn the mapping from an interaction matrix $J$ to a ground state configuration, yielding guesses for the set of most probable configurations. Given these guesses, we show that ground state configurations can be searched much faster than in vanilla simulated annealing. For large problems, a model trained on small $J$ matrices predicts a configurations whose energy is much lower than those obtained by simulated annealing, indicating the size generalizability of the strategy.
\end{abstract}

\maketitle

Finding the ground state configurations of a complex energy landscape is a long standing computational challenge \cite{wales2006potential}. Short of brute-force enumeration, random search algorithms such as simulated annealing can anneal Markov chains to the global minimum as a simulation temperature approaches zero \cite{kirkpatrick1983optimization}. However, in cases where many interacting degrees of freedom result in a highly rugged energy landscapes, conventional methods suffer from low probability of overcoming energy barriers and the chain may get stuck in local minima \cite{papadimitrou1982combinatorial, landau2014guide, WOLFF199093}.
 
The classical methods for searching energy landscapes are devised to work for general problems. Yet, many scientific problems often present themselves via an \textit{ensemble} of energy landscapes with similar underlying patterns, with interactions arising from a single or handful of governing equations. Examples include the energy landscapes of organic molecules built out of chemical building blocks, where potential energies are obtained by solving Schr\"{o}dinger equation, or the space of protein structures from interactions of individual amino acids. In an ensemble setting, we hypothesize that there exist {\sl system-specific} sampling rules \cite{wolff1989collective, wang1990cluster, houdayer2001cluster} that make it possible to traverse these particular energy landscapes more efficiently than classical methods. These rules can be learned from examples of energy minima calculated with classical methods for small problems.

Here we demonstrate this approach in the context of a model problem that defines a natural ensemble. We construct Ising spin glasses \cite{mezard1987spin, edwards1975theory}, where the interaction matrix $J$ is a \textit{structured random} matrix, chosen from protein contact maps. Given the large database of natural proteins \cite{berman2000protein} and the distinctive contact pattern of a folded protein \cite{nelson2008lehninger},  protein contact map data gives an ideal ensemble for testing whether interaction rules encoded in $J$'s are consistent across varying system sizes. 

In recent years, several machine learning techniques have been applied to sample spin configurations of Ising model. The list includes but is not limited to simple regression \cite{liu2017self}, restricted Boltzmann machine \cite{huang2017accelerated}, reinforcement learning \cite{bojesen2018policy}, autoregressive model \cite{wu2019solving, mcnaughton2020boosting}, and normalizing flow model \cite{hartnett2020self}. The goal of these works is to estimate  the Boltzmann distribution of a given problem so that the learned model can either completely replace or be used as a proposal distribution for Markov Chain Monte Carlo simulation. However, these schemes do not consider learning with $J$ instances of varying sizes.

Instead we recast spin glass energy minimization, a well known NP-hard problem \cite{barahona1982computational}, as a node classification problem in graph theory, and employ a graph neural network (GNN) \cite{wu2020comprehensive} to parametrize the mapping from a $J$ to the corresponding ground state configuration. We generate the set of most probable configurations from the GNN model to predict low-lying configurations of an energy landscape. If this configuration set misses a ground state configuration, we show that simulated annealing starting from a configuration in this set can search for the ground state configuration more efficiently. The schematic of this strategy is described in Fig.~\ref{fig:intro}. 

We further test the utility of the GNN model by constraining the size of $J$---where size refers to the number of amino acid---in a training set and testing the trained model on larger $J$'s. As we increase the size limit of training set $J$ from 30 to 500, model's test performances quickly reach the level comparable to those obtained with the size limit of 800. We also show that the model trained on $J$ with size less than 800 can predict configurations whose energies are much lower than those found by simulated annealing for $J$ with size around 3000. 

\setcounter{figure}{0}    
\begin{figure*}
  \centering
  \begin{adjustbox}{center}
    \includegraphics[width=2\columnwidth]{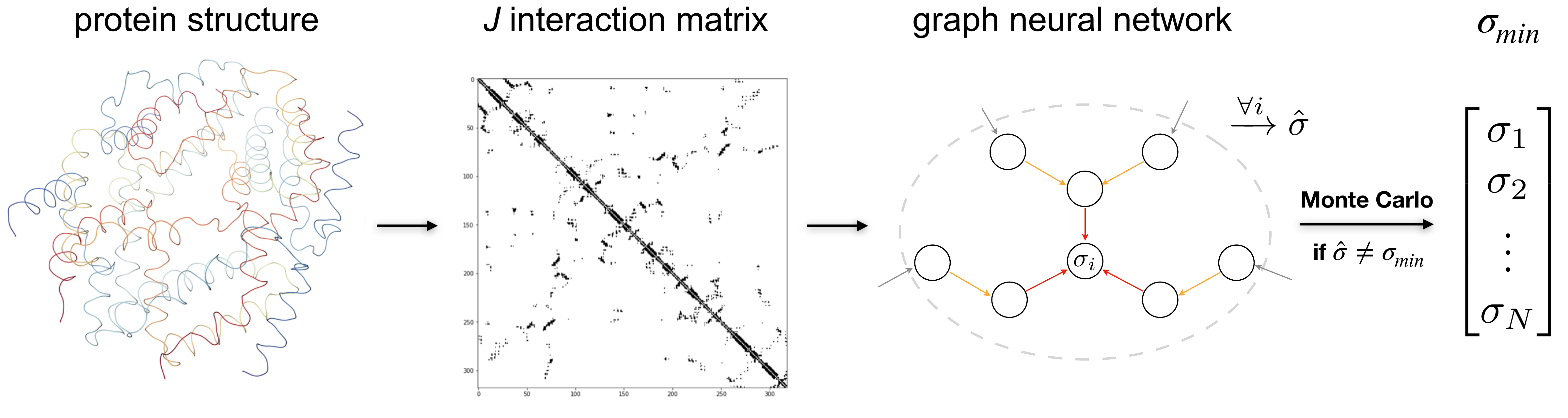}
  \end{adjustbox}
  \caption{Schematic of model formulation and ground state prediction. Binary matrices $J$'s obtained from protein structures define the set of Ising model Hamiltonian. The resulting potential energy landscapes are similar since $J$'s have the patterns of connectivity from natural protein folds. We train graph neural network with $\sigma_{min}$ found from simulated annealing. The model aggregates the nearest-neighbor information for all spins at each layer. Thus an $L$-layer model can account for $L$-hop neighborhood information. As the model learns the rule of local interaction, it predicts a configuration which, if not the ground state already, can be improved by simple configuration enumeration and Monte Carlo sampling.}
  \label{fig:intro}
\end{figure*}

We begin by constructing an ensemble of Hamiltonians for which the underlying potential energy landscapes have similar patterns.
For simplicity, we consider Ising  Hamiltonians of the form
\begin{equation}
\mathcal{H}(\sigma) = - \frac{1}{2} \sum_{i,j}^N J_{ij} \sigma_i \sigma_j + h \sum_i^N \sigma_i, \quad h = \frac{\sum J_{ij}}{2N}
\label{eq:one}
\end{equation}
where both coupling and field terms depend on an interaction matrix $J$ and $J_{ij}, \sigma_i \in \{0, 1\}$. The field is chosen to prevent all ground state configurations from  collapsing to a trivial ground state of all 1's. Within this formulation, to obtain an energy landscape ensemble, we need to specify an ensemble of $J$ matrices, whose $J$ is random yet with distinct shared patterns. Due to binary $J_{ij}$, this structured randomness of $J$ shall be encoded in spin connectivity. 

In this work, we obtain this ensemble using protein contact maps to calculate the $J$ ensemble. Since proteins are characterized by distinct secondary structures, together with non-local contacts, protein contact maps define a set of structured random connectivity matrices. We downloaded fist subunit of all protein structure files deposited in the Protein Data Bank \cite{rcsbPDB} to ensure $J$'s do not have a distinct block diagonal structure due to the presence of multiple domains. Hence the connectivity features in our $J$ ensemble solely originate from the pattern of intra-domain folding, which are referred to as {\sl secondary} and {\sl tertiary} protein structures.  We excluded protein with missing spatial information or whose protein chain is shorter than 20 residues or longer than 800 residues. We additionally added two largest subunit structures with chain length of 3661 and 2814 for the size generalizability experiments. From these files, we generated contact maps by setting $J_{ij}$ as 1 if the distance between two corresponding amino acid residues is less than 8\AA, and 0 otherwise \cite{monastyrskyy2014evaluation}. From this procedure, we obtained 64563 different contact maps, excluding the two large cases, to define our $J$ ensemble. We emphasize the resulting spin configurations derived from these energy functions Eq.~\ref{eq:one} have no relation to amino sequences; our intent here is not to make predictions about proteins {\sl per se}, but instead to use the regularity of protein structures to define a natural ensemble. 

For all $J$'s except the two largest, we ran simulated annealing for each $J$ starting from 100 random initial configurations, and selected an annealed configuration with the lowest energy as its purported ground state configuration $\sigma_{min}$. The annealing schedule was optimized such that simulated annealing always find the ground state configurations on $J$'s with size smaller than 30, which we identified from brute-force enumeration of all configurations. For the two largest $J$'s, we decreased a cooling rate and increased an equilibration steps at each temperature to account for an enlarged configurational space and ran 30 randomly initialized simulated annealing. Further simulation details are discussed in the Supplemental Material \footnote{See supplemental material at ...}. Since finding the global energy minimum of an Ising spin glass in $2^N$ configuration space is NP-hard, we settled for this repeated annealing scheme and assume $\sigma_{min}$ closely approximates the actual ground state configuration. From all pairs of $J$ and $\sigma_{min}$, 6400 pairs were randomly selected as validation set, another 6400 pairs as test set, and remaining 51763 pairs as training set. For the first size generalizability experiment, we used the same test set but sub-select from the training set the pairs whose $J$ are smaller than certain size cutoffs to make \textit{small-$J$} training sets. For the second size generalizability experiment, we used the entire training set and test on the two large $J$'s.

Our prediction task is to learn the mapping from $J{\to}\sigma_{min}$.
This can be cast as a node classification problem in graph theory and hence we parametrize the mapping with a graph neural network. Given a graph, the $L$-layer model generates an expressive feature embedding for node $\sigma_i$ by aggregating the features of all $L$-hop neighbor nodes of $\sigma_i$ as shown in Fig.~\ref{fig:intro}, and uses this embedding to classify globally whether each node shall be turned on or off. To allow generalization of the mapping across $J$'s with different size and structure, we chose a message passing framework \cite{scarselli2008graph, gilmer2017neural} with attention mechanism \cite{vaswani2017attention, velickovic2017graph}, instead of Laplacian-based convolution method \cite{defferrard2016convolutional, kipf2016semi} which requires a constant graph structure.

The inputs to the graph neural network are the adjacency matrix $J$ and node features, which are initially a node degree the field strength $h$ from Eq.~\ref{eq:one}. At each layer, the network updates node features by first applying standard nonlinear transformation---expanding feature dimension from 2 to $F$, then calculating attention coefficient $\alpha_{ij}$ to find relative importance of a neighbor node $j$ to node $i$, and taking weighted sum of neighbor nodes' features using these coefficients. To capture more information from neighbors, this process is repeated $K$ times with different set of weights and newly computed $\alpha_{ij}^k$ to produce $K{\times}F$ features for a node. The features of node $i$ are then reduced to a probability $P(\sigma_i{=}1)$ in the final layer for node classification. The functional forms of the operations are detailed in the Supplemental Material \cite{Note1}. Notably, the final model used in this work consists of six layers.

The model's predicted configuration, $\hat{\sigma}$, are then obtained by choosing the greater of the two node classification probabilities, $\texttt{argmax}[P(\sigma_i{=}0) \ P(\sigma_i{=}1)]$. However, this point estimation does not take a full advantage of the learned embedding. This scheme is especially problematic for nodes with probabilities around 0.5 because the non-argmax configurations would have been just as likely. Therefore, we generate a set of top most probable configurations from the configuration probability output of the  model, giving a broader coverage of the low-lying region in the energy landscape. To obtain $M$ such configurations, we pick top $\log_2{M}$ nodes whose $P(\sigma{=}1)$ are close to 0.5, and order all permuted configurations according to their corresponding sum combination of probabilities.

\begin{figure}
  \centering
  \begin{adjustbox}{center}
    \includegraphics[width=\columnwidth]{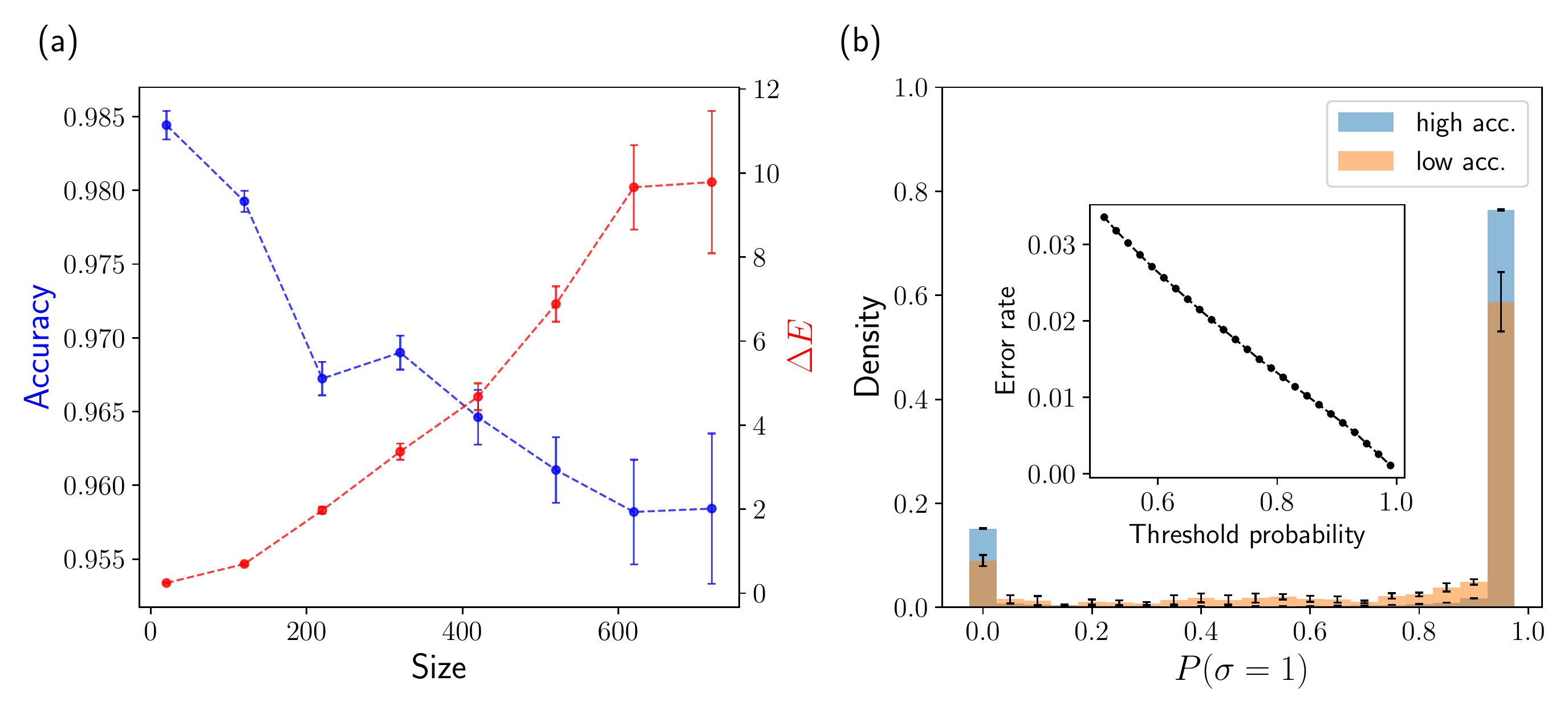}
  \end{adjustbox}
  \caption{(a) Test set performance as a function of the size of $J$, measured in accuracy (blue) and in $\Delta E$ (red). Each point in curve reports an average value for all $J$ with size window of 100. Accuracy is the fraction of correctly classified nodes and $\Delta E$ measures the difference between $E(\hat{\sigma})$ and $E(\sigma_{min})$. (b) Histogram of the classification probability for predictions whose accuracy is above 0.97 (blue) and below 0.7 (orange). Inset shows the fraction of misclassified nodes among \textit{confident} nodes as a function of the threshold probability imposed to select those nodes.}
  \label{fig:two}
\end{figure}

A GNN model trained on the entire training set correctly predicted $\sigma_{min}$ for 1700 of 6400 $J$'s in the held-out test set. To further quantify the model's performance, we investigate following two metrics. Define accuracy as the ratio between the number of correctly predicted nodes in $\hat{\sigma}$ and total number of nodes and energy difference $\Delta E$ as the energy gap between a predicted configuration and the true ground state. Fig.~\ref{fig:two}(a) shows the prediction accuracy decreases, while the energy difference increases with increasing $J$. The average accuracy and $\Delta E$ across the entire ensemble are $0.978$ and 2.79 respectively, due to the size distribution of $J$ skewed towards small $J$'s (Supplemental Material Fig.~S1 \cite{Note1}). Since energy histograms of small $J$'s obtained via complete configuration enumeration are peaked at positive energy and negative energy configurations occur in far-left tail region (Supplemental Material Fig.~S2 \cite{Note1}), predicting configurations with such small energy differences is surprising. We emphasize again that the model does not evaluate the energy function of Eq.~\ref{eq:one} to optimize a configuration. This suggests the GNN model has learned a generalizable node feature transformation for this particular class of energy landscape simply by comparing its predicted configurations to known ground state configurations.

\begin{table*}
\caption{\label{tab:table1} Summary of the size generalizability experiment on test set. Accuracy, energy offset, and the number of ground state match in 6400 test set $J$'s using the GNN's prediction $\hat{\sigma}$, the lowest-energy configuration of top most probable set $\hat{\sigma}_{top}$, and seeded annealing $\hat{\sigma}_{anneal}$ are reported.}
\begin{ruledtabular}
\begin{tabular}{cc|ccc|ccc|ccc|c}
 \multicolumn{2}{c}{Training set} &
 \multicolumn{3}{c}{$\hat{\sigma}$} & \multicolumn{3}{c}{$\hat{\sigma}_{top}$} & \multicolumn{3}{c}{$\hat{\sigma}_{anneal}$} \\
\cline{3-11}
size cutoff & \# $J$'s & $\hat{\sigma}{=}\sigma_{min}$ & acc. & $\Delta E$ & $\hat{\sigma}_{top}{=}\sigma_{min}$ & acc. & $\Delta E$ & $\hat{\sigma}_{anneal}{=}\sigma_{min}$ &  acc. & $\Delta E$ & \# $\sigma_{min}$ found \\
 \hline
30 & 560 & 14 & 0.866  & 23.84 & 150 & 0.880 & 17.22 & 2824 & 0.926 & 3.68 & 2988 \\
40 & 1301  & 65 & 0.901 & 11.05 & 260 & 0.909 & 8.68 & 2756 & 0.923 & 3.64 & 3081 \\
50 & 1839 & 388 & 0.944 & 5.15 & 1079 & 0.954 & 3.00 & 2383 & 0.932 & 2.68 & 3850 \\  
100 & 7319 & 511 & 0.953 & 4.12 & 1265 & 0.962 & 2.23 & 2272 & 0.932 & 2.74 & 4048\\
200 & 24589 & 784 & 0.964 & 3.20 & 1398 & 0.972 & 1.65 & 2222 & 0.942 & 1.85 & 4404 \\
300 & 36130  & 1442 & 0.972 & 2.79 & 1638 & 0.977 & 1.32 &  1454  & 0.938 & 1.93 & 4534\\
400 & 43190 &  1497 & 0.974 & 2.47 & 1687 & 0.980 & 1.18 & 1503  & 0.943 & 1.45 & 4687\\
500 & 47631 & 1519 & 0.976 & 2.36 & 1738 & 0.981 & 1.12 & 1463  & 0.947 & 1.28 & 4720 \\
800 & 51763 &  1700 &  0.978 & 2.31 & 1673 & 0.983 &  1.13 & 1417  & 0.953 & 1.07 & 4790\\
\end{tabular}
\end{ruledtabular}
\end{table*}

Fig.~\ref{fig:two}(b) shows the averaged histogram of node classification probability $P(\sigma{=}1)$ from high accuracy configurations in blue, and that of low accuracy configurations in orange. A striking feature is that most nodes in both cases are predicted with high certainty as evinced by the peaks at both ends. In addition, the histogram of low accuracy configurations shows more nodes in the middle, indicating that the model's prediction accuracy may be directly related to the node classification probability $P(\sigma)$. We thus set a threshold probability, $P_{thr}$, to select nodes with low uncertainty where $P(\sigma_i{=}1) \geq P_{thr}$ or $P(\sigma_i{=}1) < 1{-}P_{thr}$ and calculated an error rate among these nodes as $P_{thr}$ is varied. As shown in the inset of Fig.~\ref{fig:two}, the number of misclassified nodes among such nodes goes down as we increase the threshold. This result in turn confirms that most misclassifications indeed occur among \textit{uncertain} nodes in the middle region of the histogram. 

\begin{figure}
  \centering
  \begin{adjustbox}{center}
    \includegraphics[width=\columnwidth]{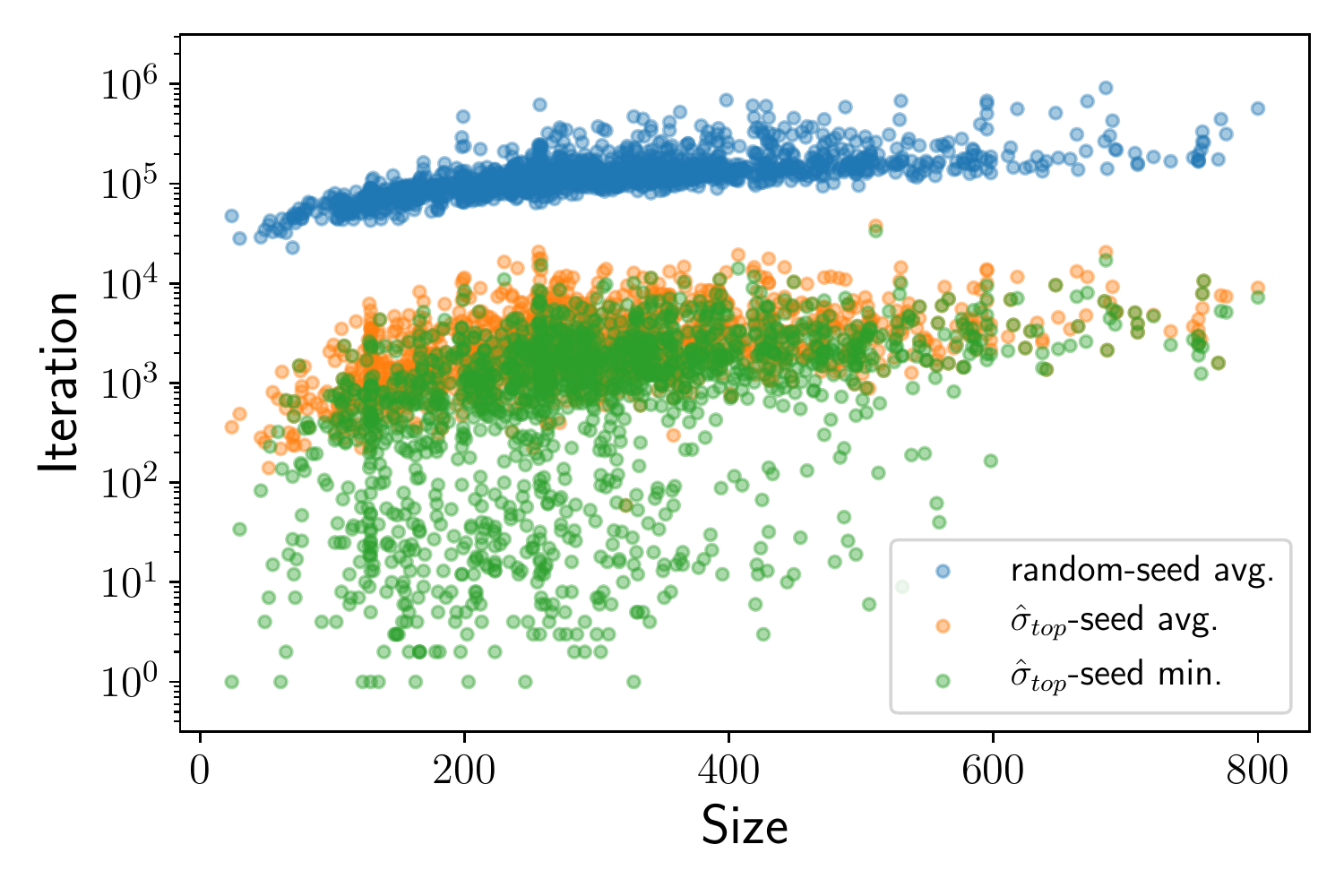}
  \end{adjustbox}
  \caption{Number of sampling steps taken to reach ground state configuration for simulated annealing launched from a random configuration (blue) and from the lowest-energy configuration of top most probable configurations (orange). Each point reports the averaged value from 10 trials for random annealing and 5 trials for seeded annealing. We also include the minimum number from the 5 trials for seeded annealing experiment (green).}
  \label{fig:three}
\end{figure}

Given there are only a handful of uncertain nodes, the set of top most probable configurations can account for most of permutations of their node configurations because the first few nodes to be changed are those with $P(\sigma_i{=}1) {\approx} 0.5$. This enumerated set allows for coverage of configuration space around the model's initial prediction. We enumerated top 1000 most probable configurations for each $J$ in the test set to cover 10 most uncertain nodes since $1000 {\approx} 2^{10}$. We then calculated the energy of these configurations and picked the lowest energy configuration as an improved prediction of the model, $\hat{\sigma}_{top}$. From this procedure, we additionally found the ground state configurations in 1673 $J$'s. This improvement of $\hat{\sigma}$ by configuration enumeration suggests that the uncertain nodes contain frustrated nodes to which the configuration energy is highly sensitive and, thus, that the GNN model has an implicit representation of energy in the node embedding.

For the half of test set where our model missed the ground state configurations, all predicted configurations still have small energy differences relative to the ground state configurations. We exploited this by running 5 simulated annealings with $\hat{\sigma}_{top}$ as a starting configuration for each remaining $J$. Since we are now annealing from a low-lying point in energy landscape, the starting temperature of the annealing should concurrently decreased to prevent the chain from sampling arbitrarily high energy states. We used the temperature value at which the energy trajectory of sampled states drifts up to a bit higher energy at the beginning to allow for initial exploration of energy landscape \cite{Note1}.  The $\hat{\sigma}_{top}$-seeded simulated annealing found the ground state configuration for additional 1417 $J$'s with about two orders of magnitude reduction in the averaged number of sampling steps as shown in Fig.~\ref{fig:three}. In about 20\% cases, the minimum number of sampling steps from 5 trials were only few hundreds as only one or two node were misclassified in $\hat{\sigma}_{top}$. In total, we found ground state configurations for 75\% of test set $J$'s. This seeded simulated annealing result shows that the predicted configuration falls in the vicinity of $\sigma_{min}$, which is often close enough that simulated annealing can locate $\sigma_{min}$. Given the top most probable configurations, we could also run simulated annealing from other configurations or perform parallel tempering \cite{swendsen1986replica, earl2005parallel} with multiple configurations to account for the possibility of $\hat{\sigma}_{top}$ falling in a basin that is too far away from the one containing $\sigma_{min}$. 

\begin{figure}
  \centering
  \begin{adjustbox}{center}
    \includegraphics[width=\columnwidth]{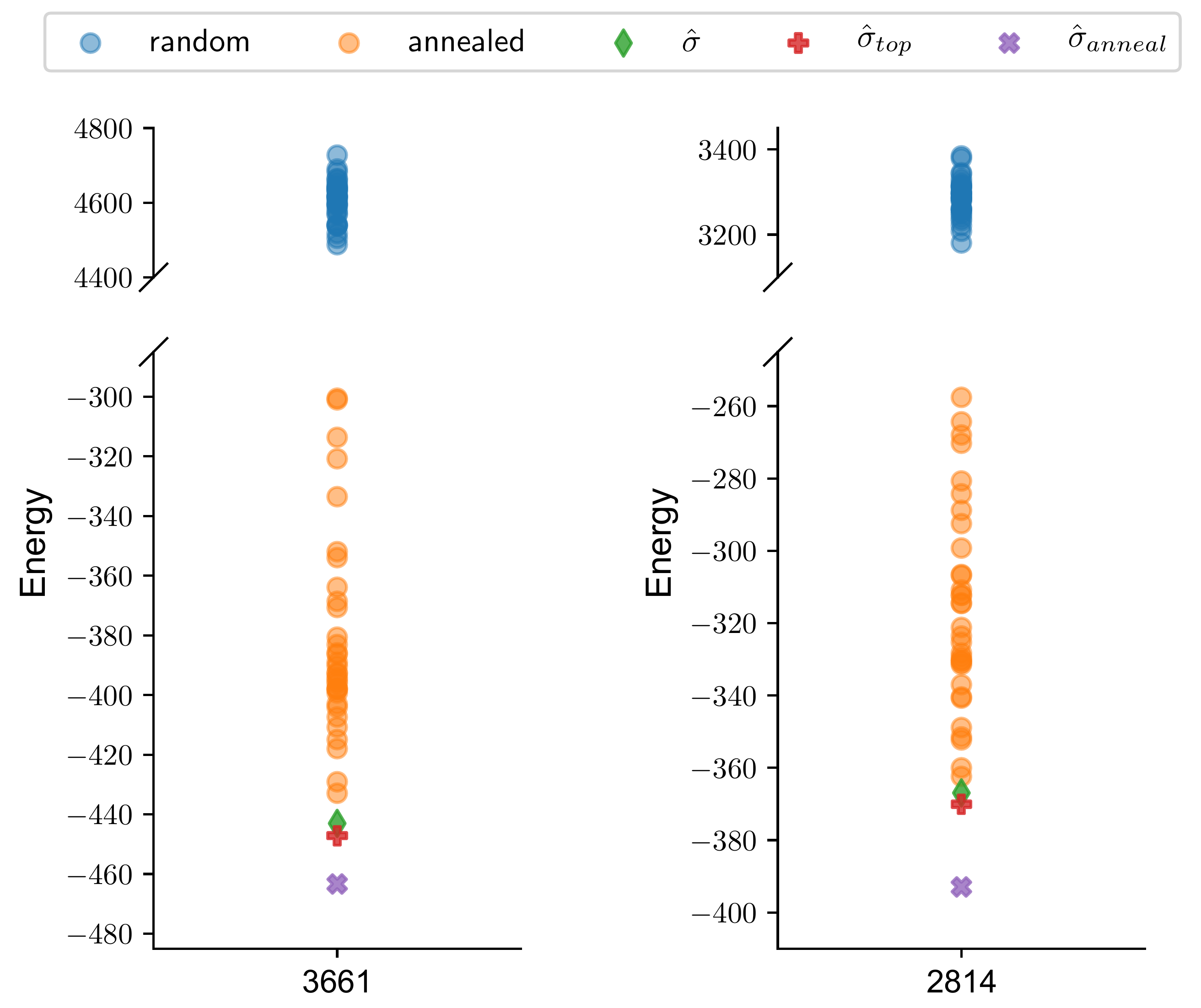}
  \end{adjustbox}
  \caption{Comparison of randomly initialized simulated annealing and the GNN model predictions on $J$ with size 3661 (left) and 2814 (right). Shown in blues are the energy of random initial configurations. Due to prohibitively large configurational space, we had to lengthen the annealing schedule to get markov chain to annealing down below -200. The GNN model's predictions in green diamond beat this dedicated effort of simulated annealing. The configuration enumeration and seed simulated annealing reach down further.}
  \label{fig:four}
\end{figure}

To test for size generalizability of GNN model, we first trained models on eight \textit{small-$J$} training sets with increasing size cutoffs and test them on the existing test set. As shown in Table~\ref{tab:table1}, test set accuracy and energy difference roughly reach those of the original model trained on the entire training set when the size cutoff for $J$ is above 300. Note that the number of the match between the annealed configurations, $\hat{\sigma}_{anneal}$, and $\sigma_{min}$ decreases as the size constraint increases because most of $\sigma$'s are already recovered through model predictions. In all cases, the configuration enumeration and seeded simulated annealing improve upon initial model predictions. If $\hat{\sigma}$, $\hat{\sigma}_{top}$, and $\hat{\sigma}_{anneal}$ are considered together, GNN model provides comparable performance even at the size cutoff of 200. The local interaction pattern of 6-hop neighbor networks in a protein shorter than 100 amino acids should be similar enough to that of much longer chain. It is thus likely that relatively poor performances with training sets with size cutoff less than 100 are simply due to a limited amount of available data. 

To test this hypothesis in more practical use case setting, we tested the original model on $J$'s of size 3661 and 2814. The model predicted $\hat{\sigma}$ with energy -443 and configuration enumeration further improved the energy to -447 for $J$ of size 3661, whereas the lowest energy configuration found from 30 randomly initialized simulated annealing runs was -432. On $J$ with size 2814, we obtained energy values of -366 for $\hat{\sigma}$ and -370 for $\hat{\sigma}_{top}$ whereas randomly initialized simulated annealing only reached down to -362. As in previous analysis, we launched simulated annealing from $\hat{\sigma}_{top}$ and obtain annealed configurations with energy -463 and -391 for $J$ with size 3661 and 2814, respectively. Fig.~\ref{fig:four} highlights the efficiency of the GNN model over randomly initialized simulated annealing.

Our work shows that it is indeed possible to use an ensemble of energy landscapes with known ground state configurations to train a neural network to deduce the ground state configuration of similar energy landscapes. On our model problem, we deterministically found the ground state configurations on 50\% of the held out test set $J$'s and stochastically on additional 25\% through a graph neural network, top configuration enumeration, and seeded simulated annealing. Although this number may appear modest, we emphasize that all configurations predicted by the model were extremely low-lying configurations, often in the vicinity of the ground state configurations. Since the loss function does not include other local minima---or, for that matter, the energy function itself, we believe that such an informed prediction is possible only if the learned node feature embedding of GNN correctly captures the local interaction rule encoded in $J$ interaction matrices, and hence the topological undulation in configuration space. 

In addition, we showcased the practical utility of the GNN model with size generalizability experiments. The GNN model predicted the configurations that could not be reached by naive simulated annealing with random initial guess, and we were able to further improve it by combining the enumeration scheme and a seeded simulated annealing. Therefore, the GNN model presents an appealing method to produce extremely good initial guesses for a class of energy landscape problem where governing physics is local. 

In future work, we will apply this framework to a variety of problems where the discovery of global minima would have technological consequences.

\begin{acknowledgments}
We thank Lucy Colwell for suggesting protein contact maps as a model system for ensembles of energy landscapes, and for helpful discussions. We also thank Yohai Bar-Sinai, Carl Goodrich, Mor Nitzan, Mobolaji Williams and Jong Yeon Lee for helpful discussions. This work is supported by the Office of Naval Research through the grant N00014-17-1-3029, as well as the Simons Foundation. 
\end{acknowledgments}

\bibliography{bibsource}

\onecolumngrid

\clearpage


\onecolumngrid
\begin{center}
\textbf{\large Supplemental Material for ``Predicting ground state configuration of energy landscape ensemble using graph neural network"}
\end{center}
\twocolumngrid

\setcounter{equation}{0}
\setcounter{figure}{0}
\setcounter{table}{0}
\setcounter{page}{1}
\makeatletter

\renewcommand{\theequation}{S\arabic{equation}}
\renewcommand{\thefigure}{S\arabic{figure}}
\renewcommand{\thetable}{S\arabic{table}}
\renewcommand{\bibnumfmt}[1]{[S#1]}
\renewcommand{\citenumfont}[1]{S#1}

\section{SIMULATION DETAILS}
For each sampling step of simulated annealing, we randomly selected a single node from a current node configuration vector $\sigma_{cur} {\in} \{0, 1\}^N$ and changed the node value to 1 if it is 0 or vice versa to generate a proposal configuration $\sigma_{prop}$, and accepted this configuration with probability $A$ given by the Metropolis criterion
\begin{equation}
    A=\text{min}\Big(1, e^{-\beta \big( \mathcal{H}(\sigma_{prop})-\mathcal{H}(\sigma_{cur}) \big)}\Big)
\end{equation}
where $\beta$ is an inverse temperature $1/T$ as we set Boltzmann constant to 1 and $\mathcal{H}$ is the Ising Hamiltonian from the main text. 

\subsection{Calibration of annealing schedule}
We used $n$-bit Gray code algorithm to enumerate all \{0,1\}-node configurations and identify ground state configurations for $J$'s with size ranging up to 30. Using an exponential cooling schedule, $T_K {=} T_0 \cdot 0.8^K$, with the initial temperature $T_0{=}10$ and exponentially increasing equilibration steps, $L_K{=}L_0 \cdot 1.2^K$, with the initial number of steps $L_0{=}1000$, 10 randomly initialized simulated annealing runs all reached those ground state configurations for $J$'s with size smaller than 30. We let the temperature cycle $K$ go up to 30. We used this annealing schedule to calculate purported ground state configuration $\sigma_{min}$ for remaining $J$'s with size ranging up to 800, repeating the simulation 100 times with different seed configurations for each $J$. Additionally, we ran several simulated annealing runs with slower cooling and longer equilibration on 50 randomly selected $J$'s with size larger than 500 but these did not improve over the lowest energy annealed configurations found from the original experiments.

\subsection{Annealing schedule for large $J$}
On the two largest $J$'s with size 3661 and 2814, all 100 simulations with the aforementioned schedule annealed to different configurations and some had relatively high energies compared to the rest, indicating a poor annealing. We thus modified the cooling schedule to $T_K {=} T_0 \cdot 0.85^K$ with the same initial temperature yet with $K$ going up to 40 and the equilibration schedule to $L_K{=}L_0 \cdot 1.15^K$ with $L_0{=}2000$. Using this schedule, we were able to anneal to low energy configurations in all 30 randomly initialized simulated annealing runs. The energies of initial configurations and annealed configurations are shown as blue and orange dots in Fig. 4 of the main text. 

\subsection{Annealing schedule for seeded simulated annealing}
Since $\sigma_{top}$ has low energy and is thus likely in the vicinity of $\sigma_{min}$, we modified the cooling schedule and equilibration length to focus our search on the configurational space near $\sigma_{top}$---i.e., \textit{exploit} rather than explore. We kept the factor of 0.8 in the exponential cooling but decreased the initial temperature $T_0$ to 0.5, which was high enough that configurations with energy higher than $\sigma_{top}$ are accepted for \textit{all} $J$'s in the test set. For the equilibration, we decreased the initial number of steps $L_0$ to 100 while keeping the exponential form of the length schedule. Although it is certainly possible to optimize the schedule per individual $J$ basis and find $\sigma_{min}$ for more $J$'s and quicker, we did not pursue this further as it deviates from the scope of this work.

\section{SIZE DISTRIBUTIONS OF $J$}
\begin{figure}[H]
  \centering
  \begin{adjustbox}{center}
    \includegraphics[width=\columnwidth]{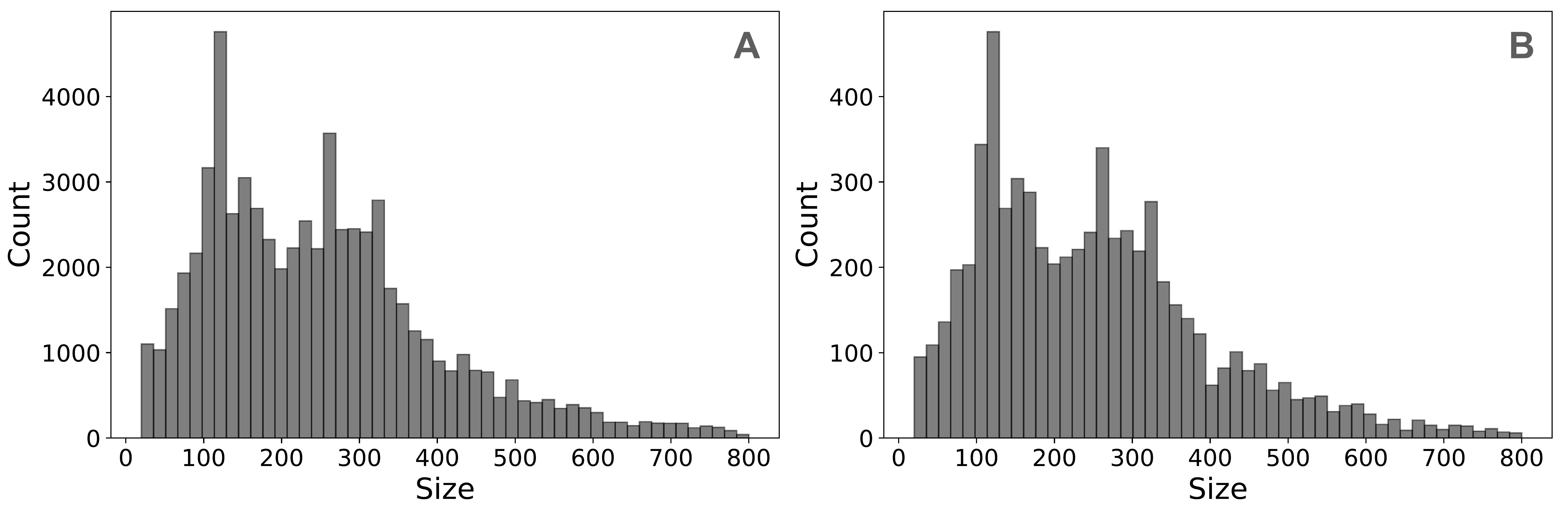}
  \end{adjustbox}
  \caption{Size histogram of entire $J$ ensemble (A) and test set $J$'s (B). They contain 64563 and 6400 $J$'s, respectively.}
  \label{fig:suppl_one}
\end{figure}

Fig.~\ref{fig:suppl_one} shows the size distributions of $J$ in the entire ensemble and test set. Note that the ensemble histogram reflects the length distribution of first subunit of all proteins in Protein Data Bank that have more than 20 but fewer than 800 amino acids. The histogram of the test set $J$ resembles the full histogram, as we would expect for a random split. 

\section{Energy histograms on small $J$}
\begin{figure}[h]
  \centering
  \begin{adjustbox}{center}
    \includegraphics[width=\columnwidth]{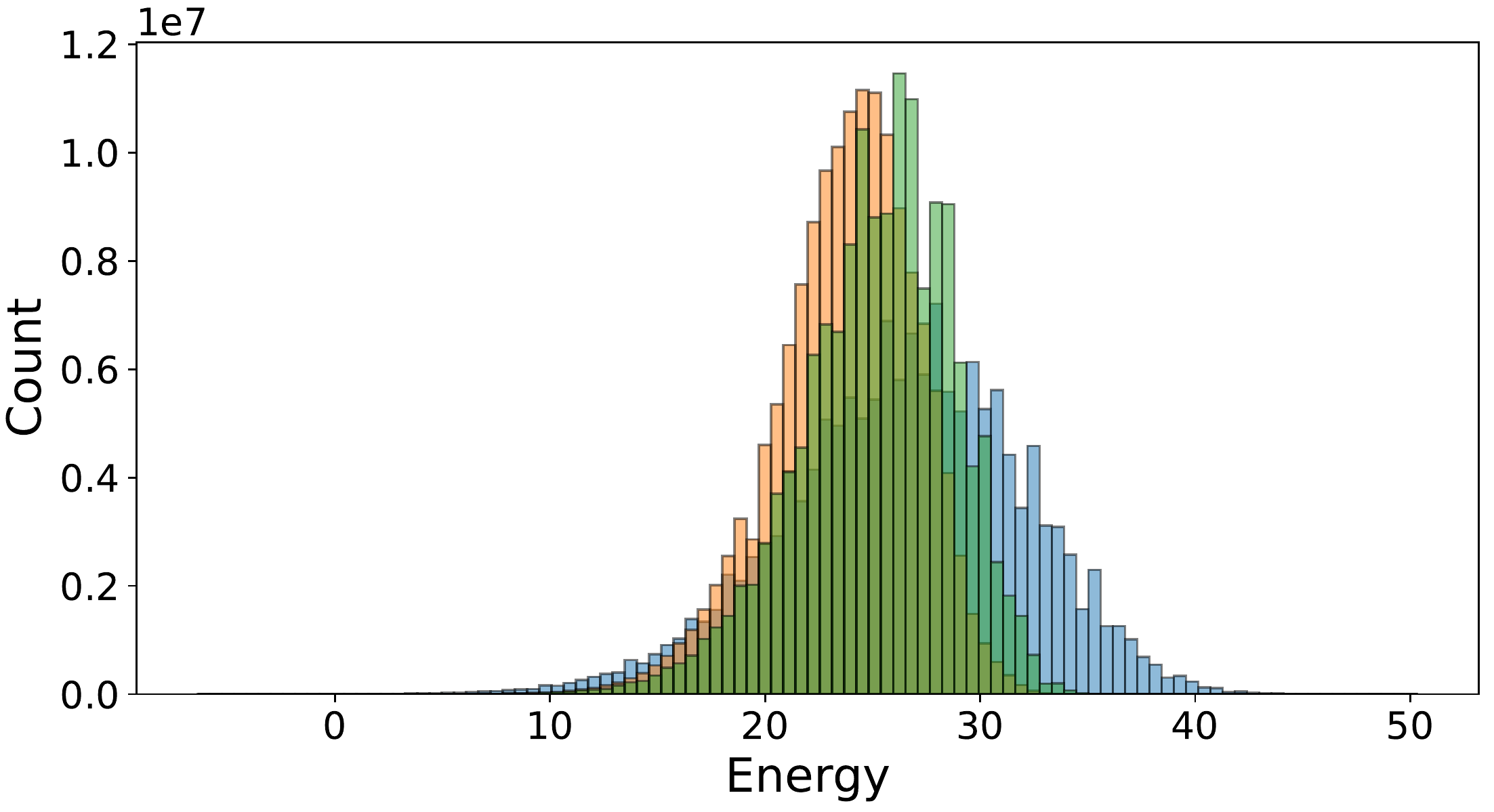}
  \end{adjustbox}
  \caption{Histograms of energy values calculated from all $2^{28}$ brute-force enumerated configurations on three $J$'s with size 28.}
  \label{fig:suppl_two}
\end{figure}

Fig.~\ref{fig:suppl_two} show the histogram of energy values of three randomly selected $J$'s with size 28, calculated using all configurations found via brute-force enumeration with 28-bit Gray code generator. This can be viewed as the density of states for the five instances of spin glass Hamiltonian of Eq.~1 of the main text. Note that negative energy values occur several standard deviations away from the center, which is at a positive energy. We expect this distributional form to hold in bigger $J$'s, yet with larger spread due to more contacts in $J$. In $J$ with size 3661, for example, the distribution seems to have a peak around 4600 and a configuration can have energy value at least as high as 4700 as shown in Fig.~4 of the main text.

\section{NETWORK STRUCTURE AND TRAINING SETUP}
The network takes two input, an adjacency matrix $J$ and node feature matrix, $\textbf{f} {=} \{\vec{f}_1, \vec{f}_2, \dots, \vec{f}_N \}$ with $\vec{f}_i {\in} \mathbb{R}^M$ where $M$ is initially 2 with node degree and field strength, $h{=}\sum J_{ij} / 2N$, as these two input node features. At each network layer, we first transform $M$ number of features into $F$ number of features via a weight matrix, $\textbf{W} {\in} \mathbb{R}^{F \times M}$. We then calculate an attention coefficient $\alpha_{ij}$---which represents the importance of node $j$ on node $i$---using a weight vector, $\vec{\textbf{a}} {\in} \mathbb{R}^{2F}$ as following
\begin{equation}
\alpha_{ij} = \frac{\exp\big(\sigma\big( \vec{\textbf{a}}^T[\mathbf{W}\vec{f}_i \| \mathbf{W}\vec{f}_j] \big)\big)}{\sum\limits_{k\in\{i,\mathcal{N}_i\}}\exp\big(\sigma\big( \vec{\textbf{a}}^T [\mathbf{W}\vec{f}_i \| \mathbf{W}\vec{f}_k] \big)\big)}
\end{equation}
where $\mathcal{N}_i$ denotes neighbors of node $i$, $\sigma$ represents an activation function, and $\|$ is concatenation operation. We used LeakyReLU activation with negative input slope $\alpha{=}0.2$ for $\sigma$. Note that the summation includes the current node $i$ to account for the influence of the node itself, relative to neighbors, on updating its features. This technique is referred to as \textit{self-attention} in the machine learning literature. The updated node feature $\vec{f}'$ is then computed by taking a weighted sum of features of neighboring nodes,
\begin{equation}
 \vec{f}'_i = \sigma \Bigg( \sum\limits_{j\in\{i,\mathcal{N}_i\}} \alpha_{ij} \mathbf{W}\vec{f}_j \Bigg)
\label{eq:suppl_three}
\end{equation}
with ELU activation function with default multiplicative factor $\alpha{=}1$ as $\sigma$.

To make node features more expressive while improving learning stability, we implemented multi-head attention where the feature update of Eq.~\ref{eq:suppl_three} is repeated for $K$ times, each with different $\textbf{W}$, $\vec{\textbf{a}}$, and newly computed $\alpha$. The $K$ sets of hidden features are concatenated in input and middle layers to generate $K \cdot M'$ features, 
\begin{equation}
\vec{f}'_i = \underset{k=1}{\overset{K}{\big\|}} \sigma \Bigg( \sum\limits_{j\in\{i,\mathcal{N}_i\}} \alpha_{ij}^k \mathbf{W}^k\vec{f}_j \Bigg)
\label{eq:suppl_four}
\end{equation}
and averaged in the final layer to retain the feature dimension $F$ which is reduced to 2, representing $P(\sigma_i{=}0)$ and $P(\sigma_i{=}1)$,
\begin{equation}
\vec{f}^{out}_i = \frac{1}{K} \sum\limits_{k=1}^K \sum\limits_{j\in\{i, \mathcal{N}_i\}} \alpha_{ij}^k \mathbf{W}^k\vec{f}_j 
\label{eq:suppl_five}
\end{equation}

We then apply sigmoid activation to the output matrix $\textbf{f}^{out}$ to obtain configuration probability matrix, $\textbf{P} {=} \{\vec{P}_1, \vec{P}_2, \dots, \vec{P}_N \}$ where $\vec{P}_i {=} [P(\sigma_i{=}0) \ P(\sigma_i{=}1)]^T$.

We trained the model to minimize the binary cross-entropy between the configuration probabilities and $\sigma_{min}$ found via simulated annealing, and optimized model hyperparameters on the 6400 validation set. We used Adam optimizer with an initial learning rate of 0.002 and a batch size of 16. The final model consisted of six graph attention layers---5 {\sl hidden} layers implementing Eq.~\ref{eq:suppl_four}, each with $K{=}4$ attention heads and $F{=}128$ node features, and a final layer implementing Eq.~\ref{eq:suppl_five}.

\section{EFFECT OF LAYER DEPTH ON TEST SET PERFORMANCE}
\begin{figure}[H]
  \centering
  \begin{adjustbox}{center}
    \includegraphics[width=\columnwidth]{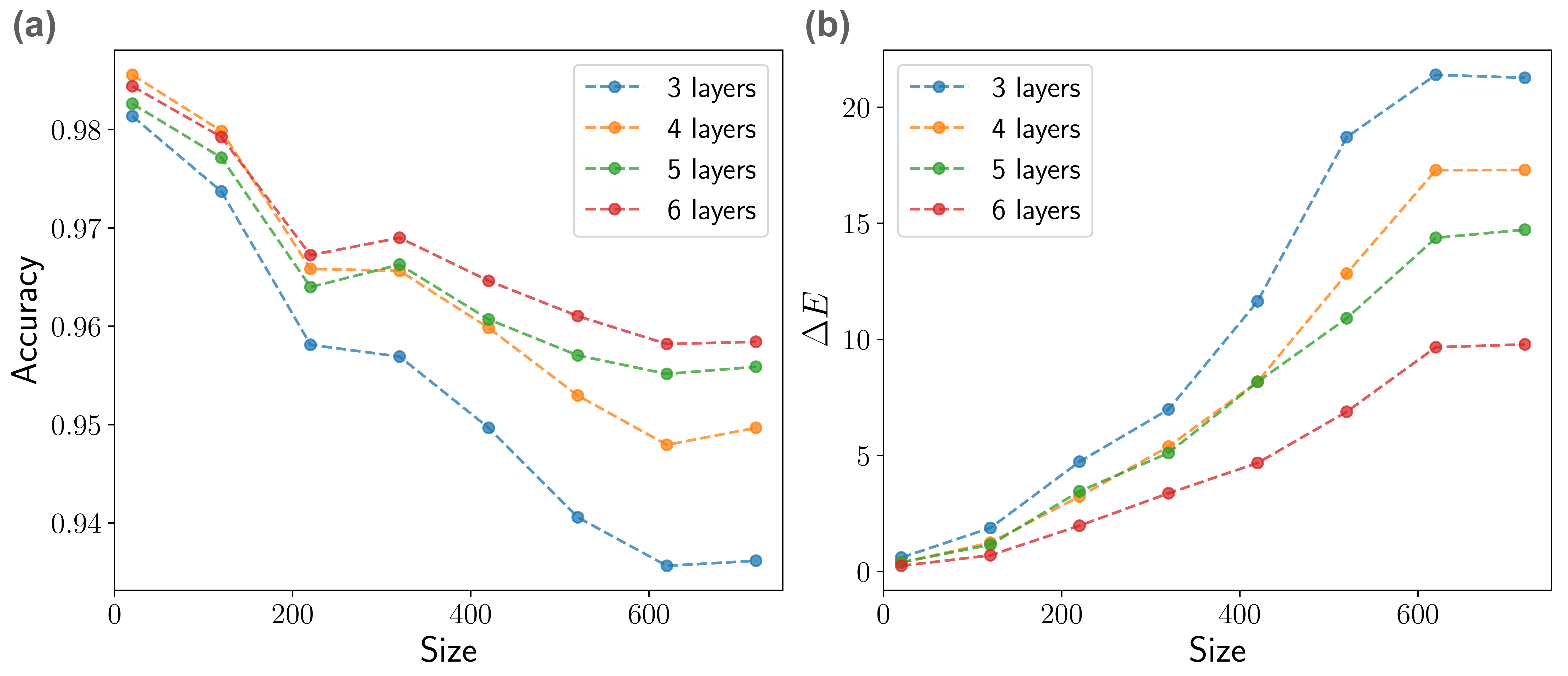}
  \end{adjustbox}
  \caption{Accuracy (a) and energy difference between $\hat{\sigma}$ and  $\sigma_{min}$ (b) over averaged size window of 100 on test set $J$ at varying graph attention layer depths. The curves for 6-layer model are identical to those shown in Fig.~2 (a) of the main text. Error bars are omitted for comparison purpose.}
  \label{fig:suppl_three}
\end{figure}

As shown in the previous section, each graph attention layer aggregates information from 1-hop neighbors. Hence, a depth of model controls the \textit{field of view} we employ in learning the local interaction rules. Fig.~\ref{fig:suppl_three} summarizes how a model's test set performance change as we increase its layer depth. We noticed that going past 6 layers give negligible improvement. From a network analysis perspective, it would be a worthwhile effort to probe whether this saturation of learnability at 6-hop neighborhood hints to a cluster property intrinsic to protein contact maps.

\section{Additional results from size generalizability experiment}
Table~\ref{tab:suppl_one} reports detailed breakdowns of the test set accuracy and the energy difference of models trained on {\sl small-$J$} training sets. \textit{over} or \textit{under} refers to a given metric averaged on all test set $J$'s whose sizes are over or under the size cutoff applied during model training. 
\begin{table}[H]
\begin{ruledtabular}
\begin{tabular}{c|cc|cc}
 \multicolumn{1}{c}{} &
 \multicolumn{2}{c}{Accuracy} & 
 \multicolumn{2}{c}{$\Delta E$} \\
\cline{2-5}
size cutoff & under & over & under & over \\
 \hline
30 & 0.881199 & 0.871548 & 2.450821 & 21.263792 \\
40 & 0.927850 & 0.899662 & 1.076943 & 11.178737 \\
50 & 0.968899 & 0.944050 & 0.228221 & 5.177260 \\
100 & 0.975691 & 0.950044 & 0.496598 & 4.471919 \\
200 & 0.973282 & 0.946137 & 0.913523 & 5.005585 \\
300 & 0.973081 & 0.955576 & 1.192279 & 4.803279 \\
400 & 0.970069 & 0.955870 & 2.192192 & 8.932947 \\
500 & 0.970202 & 0.951427 & 2.037306 & 8.844054
\end{tabular}
\caption{\label{tab:suppl_one} Test set size generalizability results with additional criteria of over or under training set size cutoffs. The number of $J$'s at each size cutoff is reported in the table of main text.}
\end{ruledtabular}
\label{table:one}
\end{table}
Fig.~\ref{eq:suppl_four} shows the two metrics at each size window of test set $J$'s, with finer size cutoff spacing from size 30 to 50. The degradation of model performance is less severe as a training set size cutoff increases. The model appears to generalize well starting from the size cutoff of 50. Note that there are only 1839 $J$'s in the training set at this cutoff, which is only about 3\% of the whole $J$ ensemble. 
\begin{figure}[H]
  \centering
  \begin{adjustbox}{center}
    \includegraphics[width=\columnwidth]{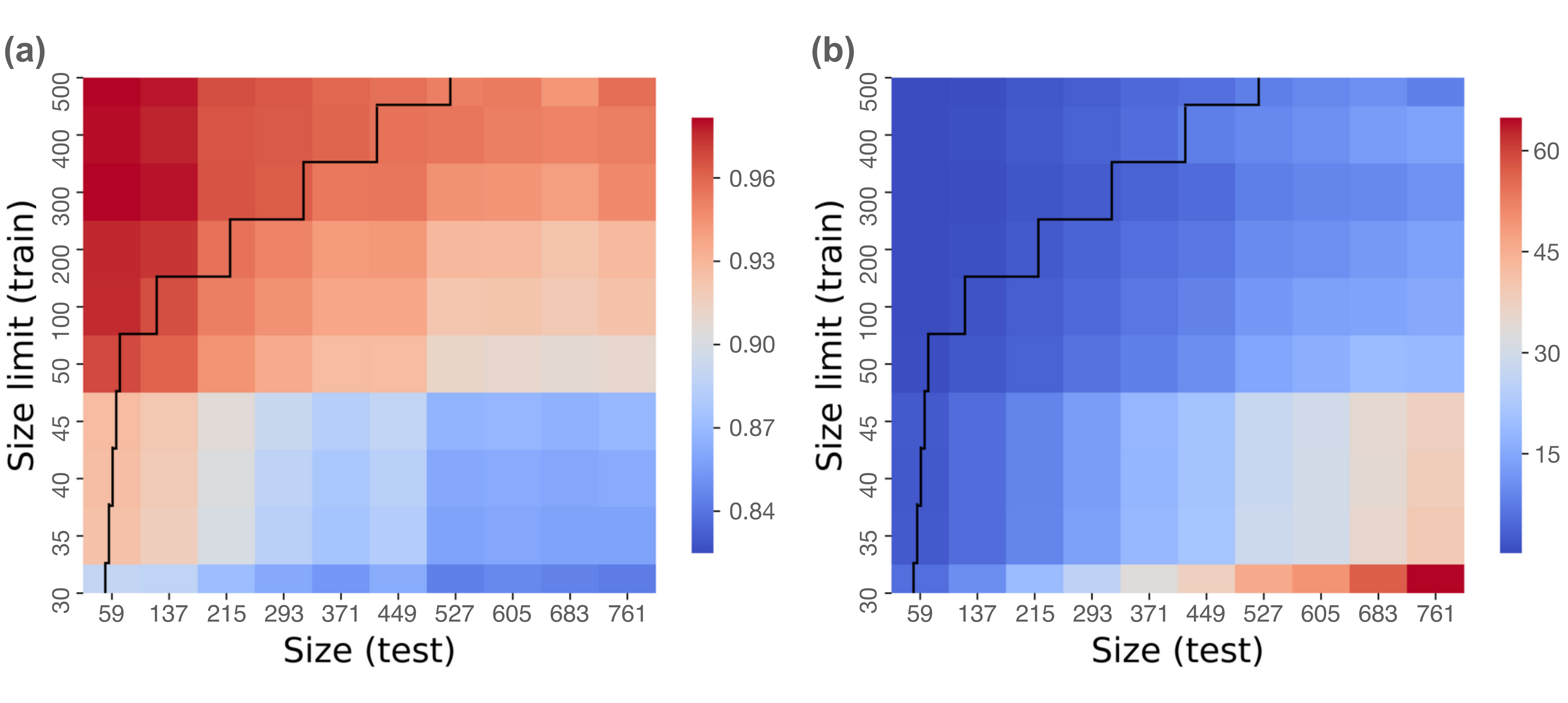}
  \end{adjustbox}
  \caption{Accuracy (a) and energy difference (b) of {\sl small-$J$}-trained models over ten size windows. Each grid reports a value averaged over all $J$'s with size $\pm$39 around size labels in the x-axis. The black line demarcates whether training set $J$'s are over or under the training set size cutoffs shown in the y-axis.}
  \label{fig:suppl_four}
\end{figure}

\end{document}